\documentclass{jpp}
\usepackage{subeqn}
\usepackage{upmath}
\usepackage{amssymb}
\usepackage{amsbsy}

\usepackage{graphicx}
\usepackage{color}   

\usepackage{natbib}
\title[Energy transfer in Hall-MHD turbulence]
      {Energy transfer in Hall-MHD turbulence: \\ 
       cascades, backscatter, and dynamo action}
\author[P.\,D. Mininni, A. Alexakis and A. Pouquet]%
       {P\ls A\ls B\ls L\ls O\ns D.\ns M\ls I\ls N\ls I\ls N\ls N\ls I%
       \thanks{e-mail: mininni@ucar.edu},\\
       A\ls L\ls E\ls X\ls A\ls N\ls D\ls R\ls O\ls S\ns 
       A\ls L\ls E\ls X\ls A\ls K\ls I\ls S%
       \thanks{e-mail: alexakis@ucar.edu}\ns \and \ns 
       A\ls N\ls N\ls I\ls C\ls K\ns P\ls O\ls U\ls Q\ls U\ls E\ls T%
       \thanks{e-mail: pouquet@ucar.edu}}
\affiliation{National Center for Atmospheric Research, 
             P.O. Box 3000, Boulder, Colorado 80307, USA}

\pubyear{2005}
\volume{0}
\part{0}
\pagerange{1--25}
\date{\today}
\begin{document}

\maketitle

\begin{abstract}
Scale interactions in Hall MHD are studied using both the mean field theory 
derivation of transport coefficients, and direct numerical simulations in 
three space dimensions. In the magnetically dominated regime, the eddy 
resistivity is found to be negative definite, leading to large scale 
instabilities. A direct cascade of the total energy is observed, although 
as the amplitude of the Hall effect is increased, backscatter of magnetic 
energy to large scales is found, a feature not present in MHD flows. 
The coupling between the magnetic and velocity fields is different 
than in the MHD case, and backscatter of energy from small scale magnetic 
fields to large scale flows is also observed. For the magnetic helicity, a 
strong quenching of its transfer is found. We also discuss non-helical 
magnetically forced Hall-MHD simulations where growth of a large scale 
magnetic field is observed. 
\end{abstract}

\section{ \label{Intro} Introduction }

The relevance of two fluid effects has recently been pointed out in 
several studies of astrophysical and laboratory plasmas 
\citep{Balbus01,Sano02,Mirnov03,Ding04}. The effect of adding the Hall 
current to the dynamics of the flow was studied in several scenarios, 
particularly dynamo action 
\citep{Helmis68,Galanti95,Mininni02,Mininni03b,Mininni05a} and 
reconnection \citep{Birn01,Shay01,Wang01,Morales05}. Several of these 
works showed that the Hall currents increase the reconnection rate 
of magnetic field lines. However, most of the studies of magnetic 
reconnection were done for particular configurations of current sheets. 
It was shown in particular by \citet{Smith04} that when a turbulent 
background is present the reconnection rate is dominated by the 
amplitude of the turbulent fluctuations. The process of 
magnetic reconnection is relevant in several astrophysical and 
geophysical scenarios, such as the magnetopause, the magnetotail, the 
solar atmosphere, or the interplanetary and interstellar medium. 
Reconnection can also play a role in the generation of large 
scale magnetic fields by dynamo action \cite{Zeldovich}.

Some of the works in Hall-magnetohydrodynamics (Hall-MHD) present 
conflicting results, indicating in some cases that the Hall effect can 
help the growth of a large scale magnetic field \citep{Mininni05a} or a 
large scale self-organization process \citep{Mahajan98,Numata04,Ohsaki05}, 
while in other cases the Hall currents were observed to generate small 
scales and filamentation \citep{Laveder02a,Laveder02b,Rheinhardt02}. 

As a result, it becomes of interest to study the physical processes 
leading to cascades and transfer of ideal invariants in three-dimensional 
Hall-MHD turbulence. Phenomena observed in the laboratory and space 
plasmas tend to show an intermittent or impulsive behavior 
\citep{Bhattacharjee99} characteristic of turbulent flows. The relevance 
of Hall-MHD turbulence in the solar wind was shown by \citet{Ghosh96}. 
Also, Hall-MHD turbulence can play a crucial role in the transfer of 
matter in the magnetopause as was pointed by \citet{Rezeau01}.

In this work, we study both analytically and numerically three dimensional 
Hall-MHD turbulence as the result of a dynamo process, and from a purely 
electromotive forcing. Detailed studies of shell-to-shell energy transfer 
from direct numerical simulations (DNS) have been done for hydrodynamic 
\citep{Domaradzki90,Ohkitani92,Zhou93,Yeung95,Alexakis05b}
and magnetohydrodynamic flows \citep{Debliquy05,Alexakis05a,Mininni05b}. 
To the best of our knowledge, the energy transfer in Hall-MHD 
turbulence has not been studied before.

We show evidence of non-locality of the transfer in Fourier space, 
and that the Hall effect can increase both the transfer of magnetic 
energy to smaller scales (locally), as well as give a novel non-local 
backscatter of magnetic energy to large scales. These results become 
clear when examining the modification to the turbulent magnetic 
diffusivity due to the Hall term. Also, we observe that the Hall 
currents impact on the coupling between the magnetic and velocity fields. 
The transfer of energy between these two fields is different than in the 
MHD case. The Hall-MHD equations also display a backscatter of energy 
from small scale magnetic fluctuations to the large scale flows themselves. 
The transfer of helicity is briefly discussed as well and observed to be 
quenched by the Hall effect.

The structure of the paper is as follows. In Sec. \ref{Equations} we 
introduce the Hall-MHD equations and we define the various transfer 
terms. In Sec. \ref{Transport} we derive turbulent transport 
coefficients for the Hall-MHD induction equation. In Sec. 
\ref{Simulations} we briefly discuss the code and details of the 
mechanically forced numerical simulations for completeness. 
Section \ref{Transfers} presents the 
transfer terms in Hall-MHD as obtained from the numerical simulations. 
Section \ref{Upscaling} shows backscatter of magnetic energy in 
non-helical Hall-MHD magnetically forced simulations. Finally, Sec. 
\ref{Discussion} summarizes the results and discusses implications of 
our work for the understanding of turbulence, dynamo action, and 
reconnection in Hall-MHD.

\section{ \label{Equations} The Hall-MHD equations and transfer terms }

In dimensionless Alfv\'enic units, the Hall-MHD equations are
\begin{equation}
\partial_t {\bf U} + {\bf U}\cdot \nabla {\bf U} = - \nabla {\cal P} + 
    {\bf B}\cdot \nabla {\bf B} + \nu \nabla^2 {\bf U} +{\bf f} ,
\label{eq:momentum}
\end{equation}
\begin{equation}
\partial_t {\bf B} = \nabla \times \left [ \left( {\bf U} - \epsilon 
    {\bf J} \right) \times {\bf B} \right] + \eta \nabla^2 {\bf B} ,
\label{eq:induction1}
\end{equation}
where ${\bf U}$ is the bulk velocity field, ${\bf B}$ is the magnetic 
field, ${\bf J} = \nabla \times {\bf B}$ is the current density, 
${\cal P}$ is the pressure, $\nu$ is the kinematic viscosity, and 
$\eta$ is the magnetic diffusivity. From the Maxwell equations and 
incompressibility of the flow, 
$\nabla \cdot {\bf U} = \nabla \cdot {\bf B} = 0$. 

The Hall term $\epsilon {\bf J} \times {\bf B}$ in Eq. (\ref{eq:induction1}) 
measures the velocity difference between species, where the electron 
velocity is ${\bf U}^e = {\bf U} - \epsilon {\bf J}$. Here, $ \epsilon $ 
measures the relative strength of the Hall effect, with the Hall term being 
dominant for wavenumbers larger than $k_{Hall} \sim 1/\epsilon$ if 
equipartition between the fields is assumed. The measure of strength 
of the Hall effect can be written as $\epsilon = L_{Hall}/L_0$ where 
$L_0$ is a characteristic length (we will use $L_0=2\pi$, the size of 
the box in our simulations). In terms of physical parameters, and for 
a fully ionized plasma, the Hall length is $L_{Hall}=cU_A/(\omega_{pi}U_0)$, 
where $U_A$ is the Alfv\'enic speed, $U_0$ is a characteristic speed, 
$c$ is the speed of light, and $\omega_{pi}$ is the ion plasma 
frequency (when $U_0=U_A$, $L_{Hall}$ reduces to the ion skin depth). 
In a partially ionized plasma, expressions for $L_{Hall}$ can be found 
in \citet{Sano02} and \citet{Mininni03b}.

Of special interest is the ratio between the integral length $L$, the 
Hall length $L_{Hall}$, and the Ohmic dissipation length $L_\eta$. For 
$L_{Hall} \ll L_\eta$ ($\epsilon \to 0$), the Hall-MHD equations reduce 
to the well known MHD case. In several astrophysical problems, such as 
accretion disks, protoplanetary disks, or the magnetopause 
\citep[see e.g.][]{Birn01,Balbus01,Sano02}, the Hall scale is larger than 
Ohmic scales although smaller than the integral scale $L$ of the flow. 
We will be interested in this regime in this work, although we remark 
that the separation between these scales in astrophysical or geophysical 
problems is far from what can be achieved in numerical simulations.

The Hall-MHD equations have three ideal invariants \citep{Turner86}. In 
this work we will focus on two invariants, the total energy
\begin{equation}
E = \frac{1}{2} \int (U^2 + B^2) \, d{\bf x}^3 ,
\end{equation}
and the magnetic helicity
\begin{equation}
H = \frac{1}{2} \int {\bf A} \cdot {\bf B} \, d{\bf x}^3 ,
\end{equation}
where {\bf A} is the vector potential, $\nabla \times {\bf A} =  {\bf B}$. 
These quantities are also ideal invariants of the MHD equations 
($\epsilon=0$). The third MHD invariant, the cross helicity, is replaced 
in Hall-MHD by the hybrid helicity \citep{Turner86} 
and is small in the simulations we will discuss.

The expressions we will use for the shell-to-shell energy transfers have
been derived for the MHD case by \citet{Verma04,Debliquy05}; and 
\citet{Alexakis05a}. Here we present the derivation of 
the transfer terms for the Hall-MHD equations. Equation 
(\ref{eq:induction1}) can be rewritten as
\begin{equation}
\partial_t {\bf B} + {\bf U}\cdot \nabla {\bf B} = {\bf B}\cdot \nabla 
    {\bf U} - \epsilon \nabla \times \left( {\bf J} \times {\bf B} \right) 
    + \eta \nabla^2 {\bf B} .
\label{eq:induction}
\end{equation}
We introduce a filter in shells in Fourier space, such as ${\bf F}_K$ which
denotes the components of the field with wavenumbers between $K$ and 
$K+1$ [i.e. 
$ {\bf F}_K ({\bf x})= \sum_{k=K}^{K+1} \hat{\bf F}({\bf k}) e^{i {\bf k} 
\cdot {\bf x}} $], from Eqs. (\ref{eq:momentum}) and (\ref{eq:induction}) 
we can write detailed balance equations for the energy,
{\setlength\arraycolsep{2pt}
\begin{eqnarray}
\partial_t E_U(K) &=& \int \left\{ \sum_Q \left[ - {\bf U}_K 
   {\bf \cdot (U \cdot \nabla) \cdot U}_Q + {\bf U}_K 
   {\bf  \cdot (B \cdot \nabla) \cdot B}_Q \right] - \right. {} 
 \nonumber \\
&&{} - \nu \nabla^2 {\bf U}_K + {\bf f \cdot U}_K \Bigg\} \; d{\bf x}^3  \, ,
\end{eqnarray}}
{\setlength\arraycolsep{2pt}
\begin{eqnarray}
\partial_t E_B(K) &=& \int \left\{ \sum_Q \left[ - {\bf B}_K 
   {\bf \cdot (U \cdot \nabla) \cdot B}_Q + {\bf B}_K 
   {\bf \cdot (B \cdot \nabla) \cdot U}_Q + {} \right. \right.
 \nonumber \\
&&{} + \epsilon {\bf J}_K \cdot ({\bf B} \times {\bf J}_Q) \Big] 
   - \eta \nabla^2 {\bf B_K} \Bigg\} \; d{\bf x}^3 \, .
\end{eqnarray}}
Here, $E_U(K)$ and $E_B(K)$ denote respectively the kinetic and magnetic 
energy in the shell $K$. The above equations can be written in the more 
compact form:
\begin{equation}
\partial_t E_U(K) = \sum_Q [{\mathcal T}_{UU}(K,Q)+{\mathcal T}_{BU}(K,Q)] 
    - \nu {\mathcal D}_U(K) + {\mathcal F}(K) ,
\label{eq:Eu}
\end{equation}
\begin{equation}
\partial_t E_B(K) = 
\sum_Q [{\mathcal T}_{UB}(K,Q)+{\mathcal T}_{BB}(K,Q)] - 
\eta {\mathcal D}_B(K).
\label{eq:Eb}
\end{equation}
The functions ${\mathcal T}_{UU}(K,Q)$, 
${\mathcal T}_{UB}(K,Q)$, ${\mathcal T}_{BB}(K,Q)$, and 
${\mathcal T}_{BU}(K,Q)$ 
express the energy transfer between different fields and shells,
\begin{equation}
{\mathcal T}_{UU}(K,Q) \equiv 
    -\int {{\bf U}_K ({\bf U} \cdot \nabla) {\bf U}_Q} \; d{\bf x}^3 ,
\label{eq:TUU}
\end{equation}
\begin{equation}
{\mathcal T}_{UB}(K,Q) \equiv
\int {{\bf U}_K ({\bf B} \cdot \nabla) {\bf B}_Q} \; d{\bf x}^3 ,
\label{eq:TUB}
\end{equation}
\begin{equation}
{\mathcal T}_{BU}(K,Q) \equiv
\int {{\bf B}_K ({\bf B} \cdot \nabla) {\bf U}_Q} \; d{\bf x}^3 .
\label{eq:TBU}
\end{equation}
In general, for positive transfer, the first subindex denotes the 
field that receives energy, the second subindex the field that gives 
energy. The first wavenumber corresponds to the field receiving 
energy, and the second wavenumber to the field giving energy. As 
an example, positive ${\mathcal T}_{UU}(K,Q)$ represents energy 
transfered from the velocity field at the shell $K$ to velocity 
field at the shell $Q$. In the same way, positive $T_{UB}(K,Q)$ 
represents energy transfered from the magnetic field at wavenumbers 
$K$ to the velocity field at wavenumbers $Q$.

The transfer of magnetic to magnetic energy ${\mathcal T}_{BB}(K,Q)$ 
in Hall-MHD consists of two terms
\begin{equation}
{\mathcal T}_{BB}(K,Q) = {\mathcal T}_{BB}^\textrm{MHD}(K,Q) + 
    {\mathcal T}_{BB}^\textrm{Hall}(K,Q)
\end{equation}
where
\begin{equation}
{\mathcal T}_{BB}^\textrm{MHD}(K,Q) \equiv
    -\int {{\bf B}_K ({\bf U} \cdot \nabla) {\bf B}_Q  } \; d{\bf x}^3, 
\label{eq:TBBMHD}
\end{equation}
is the usual MHD transfer of magnetic energy through advection by the 
bulk velocity field, and 
\begin{equation}
{\mathcal T}_{BB}^\textrm{Hall}(K,Q) \equiv
    \epsilon \int {{\bf J}_K \cdot ({\bf B} \times {\bf J}_Q)} \; d{\bf x}^3, 
\label{eq:TBBHall}
\end{equation}
is the transfer of magnetic energy due to the Hall current. Note that 
the definition of the transfer terms corresponds to the MHD case in 
\citet{Alexakis05a}, except for the new term 
${\mathcal T}_{BB}^\textrm{Hall}(K,Q)$. However, as will be shown later, 
the behavior of the rest of the transfer terms in Hall-MHD will also be 
indirectly modified by the presence of the Hall effect.

All these transfer functions satisfy the identity
\begin{equation}
\label{tran_id}
{\mathcal T}_{vw}(K,Q)=-{\mathcal T}_{wv}(K,Q) ,
\end{equation}
where $v,w$ can be either $U$ or $B$. This detailed conservation is 
what allows us to define the terms as transfers of energy between shells. 
Note that other groupings of the nonlinear terms in the Hall-MHD equations 
would not satisfy this symmetry condition.

In Eqs. (\ref{eq:Eu}-\ref{eq:Eb}) we also have two dissipation functions 
and the energy injection rate
\begin{equation}
\nu {\mathcal D}_{U}(K) \equiv  \nu   \int |{\bf \nabla U}_K|^2  d{\bf x}^3 ,
\end{equation}
\begin{equation}
\eta {\mathcal D}_{B}(K) \equiv  \eta \int |{\bf \nabla B}_K|^2 d{\bf x}^3 ,
\end{equation}
\begin{equation}
{\mathcal F}(K) \equiv \int { \bf f} \cdot {\bf U}_K  \, d{\bf x}^3 .
\end{equation}

Finally, we can also define the transfer of magnetic helicity. From Eq. 
(\ref{eq:induction1}) we have
\begin{equation}
\partial_t H(K) = \sum_Q {\mathcal T}_H (K,Q) - \eta {\mathcal D}_H (K),
\end{equation}
where the transfer of magnetic helicity from the wavenumber $K$ to the 
wavenumber $Q$ is given by
\begin{equation}
{\mathcal T}_H (K,Q) \equiv \int{ {\bf B}_K \cdot \left[\left({\bf U} 
    - \epsilon {\bf J}\right) \times {\bf B}_Q \right] d{\bf x}^3}.
\label{eq:TH}
\end{equation}
This transfer function satisfies the relation 
${\mathcal T}_H (K,Q)=-{\mathcal T}_H (K,Q)$. 
The first term in Eq. (\ref{eq:TH}) proportional to 
${\bf U}$ is the usual transfer of $H_M$ in MHD, while the second term 
proportional to $\epsilon {\bf J}$ is the contribution due to the Hall 
effect. Note that as a whole, magnetic helicity is transfered between 
the shells $K$ and $Q$ interacting with the electron velocity field 
${\bf U} - \epsilon {\bf J}$. This is in agreement with the fact that 
in the ideal limit the magnetic field in the Hall-MHD system is frozen 
to the electron velocity field, instead of the bulk velocity field of 
the plasma as in MHD.

The dissipation rate of magnetic helicity at the wavenumber $K$ is given by
\begin{equation}
\eta {\mathcal D}_H (K) \equiv \eta \int{{\bf B}_K \cdot {\bf J}_K d{\bf x}^3}.
\end{equation}
It is also worth noting that, since the magnetic helicity is not a positive 
defined quantity contrary to the energy, the interpretation of its transfer 
is more difficult. We will not attempt here a separation of its different sign 
components \citep[see e.g.][]{Waleffe91,Chen03a,Chen03b} for the case of 
kinetic helicity in hydrodynamic turbulence).

\section{ \label{Transport} Transport coefficients }

In \citet{Mininni02}, the expression of the $\alpha$ dynamo coefficient 
was derived for Hall-MHD. Although an expression of the Hall-MHD turbulent 
diffusivity was derived by \citet{Mininni03a}, the closure was 
only valid for specific solutions of the Hall-MHD equations. To 
interpret the results from the energy transfer, it will be useful to 
have expressions for all the turbulent transport coefficients in the 
induction equation. To this end, and for the sake of simplicity, we 
will use mean field theory (MFT) \citep{Steenbeck66,Krause} and the 
reduced smooth approximation (RSA) \citep{Blackman99}. RSA was introduced 
to solve some ambiguities present in MFT when the magnetic field is 
strong enough to affect the velocity field trough the Lorentz force. 
Although there are still assumptions in MFT not completely justified, 
at least a qualitative agreement has been observed with simulations 
in the MHD case \citep{Brandenburg01} and the Hall-MHD case 
\citep{Mininni03b,Mininni05a}. The transport coefficients can also be 
derived using more elaborate closures, such as the Lagrangian History 
Direct Interaction Approximation (LHDIA) or the Eddy Damped Quasi Normal 
Markovian (EDQNM) closures \citep[see e.g.][]{Lesieur}. It is worth noting 
that the analysis that follows in Sections \ref{Simulations} and 
\ref{Transfers} is of general validity and independent of the assumptions 
we will use here to derive the turbulent transport coefficients.

We split the fields into
\begin{eqnarray}
{\bf U} &=& \overline{\bf U} + {\bf u} + {\bf u}_0 , \\
{\bf B} &=& \overline{\bf B} + {\bf b} + {\bf b}_0 ,
\end{eqnarray}
where ${\bf u}_0$ and ${\bf b}_0$ are isotropic and homogeneous 
solutions of Eqs. (\ref{eq:momentum}) and (\ref{eq:induction}) in 
the absence of the mean fields $\overline{\bf U}$ and $\overline{\bf B}$. 
The fields with overbars are large scale fields, and ${\bf u}$ and ${\bf b}$ 
are small scale corrections to the isotropic and homogeneous solutions 
due to the presence of the large scale fields. The fluctuating fields 
satisfy 
$\left< {\bf u} \right> = \left< {\bf u}_0 \right> = 
    \left< {\bf b} \right> = \left< {\bf b}_0 \right> = 0 $, 
where the brackets denote an average that satisfies Taylor's 
hypothesis \citep{Krause}. Replacing in Eq. (\ref{eq:induction}), using 
the equations for the ${\bf u}_0$ and ${\bf b}_0$ fields, dropping terms 
quadratic in the fluctuating fields ${\bf u}$ and ${\bf b}$, and averaging leads to
\begin{equation}
\partial_t \overline{\bf B} = \nabla \times \left[ \left( \overline{\bf U} 
    - \epsilon \overline{\bf J} \right) \times \overline{\bf B} 
    + \varepsilon \right] + \eta \nabla^2 \overline{\bf B} ,
\label{eq:largeB}
\end{equation}
where $\varepsilon$ is the mean field electromotive force
\begin{equation}
\varepsilon = \left< {\bf u}_0^e \times {\bf b} + {\bf u}^e \times 
    {\bf b}_0 \right> .
\label{eq:emfepsilon}
\end{equation}
Our main aim in this section is to close Eq. (\ref{eq:largeB}) and 
write $\varepsilon$ only as a function of averages of the fields 
${\bf u}_0$, ${\bf b}_0$, and spatial derivatives of $\overline{\bf B}$. 
A simple argument of symmetry shows that in the approximately isotropic 
case
\begin{equation}
\varepsilon = \alpha \overline{\bf B} - \beta \nabla \times \overline{\bf B} 
    + \gamma \nabla \times \nabla \times \overline{\bf B} .
\label{eq:emf}
\end{equation}

From Eqs. (\ref{eq:momentum}) and (\ref{eq:induction}), and subtracting 
the equations for the mean flows, we can also write equations for the 
evolution of the turbulent fluctuations ${\bf u}$ and ${\bf b}$. We drop 
terms quadratic in ${\bf u}$ and ${\bf b}$, and keep only terms to 
zeroth and linear order in $\overline{\bf B}$,
\begin{equation}
\partial_t {\bf b} = \nabla \times \left( \overline{\bf U} \times {\bf b}_0 
    + {\bf u}_0^e \times \overline{\bf B} + \epsilon {\bf b}_0^e \times 
    \overline{\bf J} + {\bf u}^e \times {\bf b}_0 + 
    {\bf u}^e_0 \times {\bf b} - \varepsilon \right) + 
    \eta \nabla^2 {\bf b} .
    \label{eq:smallb}
\end{equation}
In this equation, $\varepsilon$ involves averaged quantities, and 
from the Taylor's hypothesis it gives no contribution to the mean 
electromotive force. The fourth and fifth terms on the {\it r.h.s.} can be 
dropped using RSA, namely that $|{\bf u}| , |{\bf b}| \ll |\overline{\bf B}|$ 
(note that this condition is less stringent than the usual assumptions 
in MFT, since the amplitude of the fields ${\bf u}_0$, ${\bf b}_0$ can 
be much larger than the amplitude of the mean magnetic field). We will 
assume the viscosity and diffusivity are small, and as a result we will 
also drop the last term on the right hand side of Eq. (\ref{eq:smallb}). 
Terms proportional to $\overline{\bf U}$ can be removed in the proper 
frame of reference. Finally we obtain
\begin{equation}
\partial_t {\bf b} \approx \nabla \times \left( {\bf u}_0^e \times 
    \overline{\bf B} + \epsilon {\bf b}_0^e \times \overline{\bf J} \right) .
    \label{eq:RSAb}
\end{equation}
The second term on the right hand side of Eq. (\ref{eq:RSAb}) involves 
only spatial derivatives of $\overline{\bf B}$ and gives no contribution 
to the $\alpha$ coefficient, but is retained here since it will give 
contributions to $\beta$ and $\gamma$.

Following the same steps, we can also write an equation for the evolution 
of ${\bf u}$,
\begin{equation}
\partial_t {\bf u} \approx \overline{\bf B} \cdot \nabla {\bf b}_0 
    + {\bf b}_0 \cdot \nabla \overline{\bf B} - \nabla p .
    \label{eq:smallu}
\end{equation}

To obtain the mean field electromotive force we replace time derivatives 
in Eqs. (\ref{eq:smallb}) and (\ref{eq:smallu}) by the inverse of a 
correlation time $\tau$. This step, common in MFT, assumes the existence 
of a finite correlation time. At present there is no evidence of its 
validity in general
\citep[see e.g.][]{Gruzinov95,Blackman02,Brandenburg05}. Since we are 
introducing a correlation time to close these equations, the expressions 
obtained for the turbulent transport coefficients will be considered as 
symbolic expressions.

Before replacing the expression for ${\bf u}$ in 
Eq. (\ref{eq:emf}), Eq. (\ref{eq:smallu}) has to be solved for the small 
scale pressure $p$. We will use a technique developed by \citet{Gruzinov95} 
\citep[see also][]{Blackman02}. The $\alpha$ effect is linear in 
$\overline{\bf B}$ and therefore the correct result can be obtained 
assuming $\overline{\bf B}$ uniform. Then 
$\partial_t {\bf u} \approx \overline{\bf B} \cdot \nabla {\bf b}_0 $. 
Replacing the time derivative by $\tau^{-1}$ and replacing the expression 
in Eq. (\ref{eq:emf}), we obtain in the weak isotropic case
\begin{equation}
\alpha = \frac{\tau}{3} \left< -{\bf u}_0^e \cdot \nabla \times 
    {\bf u}_0^e + {\bf b}_0 \cdot \nabla \times {\bf b}_0 - 
    \epsilon {\bf b}_0 \cdot \nabla \times \nabla \times  
    {\bf u}_0^e \right> .
\end{equation}

To compute $\beta$ and $\gamma$ we have to keep spatial derivatives of 
$\overline{\bf B}$ in Eq. (\ref{eq:smallu}), and therefore we have to 
solve for the pressure. This was done by \citet{Gruzinov95} transforming 
Eq. (\ref{eq:smallu}) to Fourier space, and doing a Taylor expansion of 
the projector operator for incompressible ${\bf u}$ assuming a large 
scale separation between the mean and fluctuating fields. In three spatial 
dimensions, it was shown that the pressure and 
${\bf b}_0 \cdot \nabla \overline{\bf B}$ terms give no contribution to 
$\varepsilon$ in Eq. (\ref{eq:emfepsilon}). As a result, we are 
only left with the terms proportional to spatial derivatives of 
$\overline{\bf B}$ when Eq. (\ref{eq:RSAb}) and the first term on the 
r.h.s. of Eq. (\ref{eq:smallu}) are replaced on Eq. (\ref{eq:emfepsilon}). 
Again, assuming weak isotropy, we obtain the expressions 
for the remaining turbulent transport coefficients
{\setlength\arraycolsep{2pt}
\begin{eqnarray}
\beta &=& \frac{\tau}{3} \left< {{\bf u}_0^e}^2 + \epsilon \left( {\bf u}_0 
    \cdot \nabla \times {\bf b}_0^e + {\bf b}_0 
    \cdot \nabla \times {\bf u}_0^e \right) \right> , \\
\gamma &=& - \frac{\tau \epsilon}{3} \left< {\bf b}_0 \cdot {\bf u}_0^e 
    \right> .
\end{eqnarray}}
The two last terms in $\alpha$, and the third term in $\beta$ come from 
the small scale momentum equation and are related with the backreaction 
of the magnetic field into the velocity field. In the kinematic regime 
of a dynamo, \
$\alpha = - \tau/3\left<{\bf u}_0^e\cdot\nabla\times {\bf u}_0^e\right>$, 
which for $\epsilon=0$ reduces to the MHD case \citep{Krause}. The general 
expression for $\epsilon=0$ reduces to the MHD expression first found 
using the EDQNM closure by \citet{Pouquet76}, 
$\alpha = \tau/3\left<-{\bf u}_0^e\cdot\nabla\times {\bf u}_0^e 
    + {\bf b}_0\cdot\nabla\times {\bf b}_0 \right>$. Note also that in 
the MHD case in three dimensions, the turbulent diffusivity 
$\beta=-\tau \left< {{\bf u}_0}^2 \right>/3 $ is not changed during the 
nonlinear saturation \citep{Gruzinov95}.

The turbulent diffusivity $\beta$ in Hall-MHD is not positive definite, 
contrary to the pure MHD case \citep[note that negative effective 
diffusivities can be found in MHD if the assumption of homogeneity 
is dropped, see e.g.][]{Lanotte99}. A negative value of $\beta$ represents 
non-local transfer of energy from the small scale turbulent fields to the 
large scale magnetic field. This result will be of interest in the 
following sections.

It is worth studying the values of $\beta$ for particular cases. If 
$\epsilon$ is large enough and the system is magnetically dominated 
($E_B \gg E_U$), then 
$\beta \approx - \tau \epsilon^2 \left< j_0^2 \right>/3$, where 
${\bf j}_0 = \nabla \times {\bf b}_0$ and we assumed the average is 
a spatial average. In this case, $\beta$ is always negative
implying transfer of energy from the small scales to the large.

The normal modes of the Hall-MHD equations are circular polarized 
($\nabla \times {\bf u}_0^e = \pm k {\bf u}_0^e$, 
$\nabla \times {\bf b}_0 = \pm k {\bf b}_0$) and dispersive, and in the 
limit $k\gg1$ they satisfy dispersion relations 
$\omega \sim \epsilon k^2 \overline{B}$ (whistlers, right-handed polarized) 
and $\omega \sim \overline{B}/ \epsilon$ (ion-cyclotron waves, left-handed 
polarized). Also, for these waves the fields are related by 
${\bf b}_0 = -k \overline{B}/\omega {\bf u}_0^e$ \citep{Mininni05c}. 
If we assume a background of waves, 
$\beta \sim \tau/3 \left< {u_0^e}^2 \right> (1 \pm 2 \epsilon k^2 
    \overline{B} / \omega)$, which for $k$ large enough can give positive 
or negative turbulent diffusivity according to the orientation of the wave. 
Note that from the dispersion relations, at small scales whistlers give a 
finite contribution to the turbulent diffusivity, while ion-cyclotron waves 
give a much larger turbulent diffusivity that grows as $k^2$.

\section{ \label{Simulations} Simulations }

In this section we summarize the simulations that will be used to 
compute the energy and helicity transfer functions defined in 
Sec. \ref{Equations}. We performed three simulations in three 
dimensions with periodic boundary conditions, using a pseudospectral 
Hall-MHD code as described in \citet{Mininni03b,Mininni05a}. Runge-Kutta 
of second order is used to evolve the system of Eqs. (\ref{eq:momentum}) 
and (\ref{eq:induction1}). To ensure the divergence-free condition for the 
magnetic field, a curl is removed from Eq. (\ref{eq:induction1}) and the 
equation for the vector potential is instead solved, with the Couloumb's 
gauge $\nabla \cdot {\bf A} = 0$. The three simulations are done with a 
spatial resolution of $N^3=256^3$ grid points. The $2/3$ dealiasing rule is 
used, and as a result the maximum wavenumber resolved by the code is 
$k_{max} = N/3 \approx 85$. The kinematic viscosity and magnetic 
diffusivity are set to $\nu = \eta = 2 \times 10^{-3}$, and all the 
simulations are well resolved, in the sense that the kinetic 
[$k_\nu = (\left<\omega^2\right>/\nu^2)^{1/4}$] and magnetic 
[$k_\eta = (\left<J^2\right>/\eta^2)^{1/4}$] dissipation wavenumbers 
are smaller than $k_{max}$ at all times.

\begin{figure}
\begin{center} \includegraphics[width=9cm]{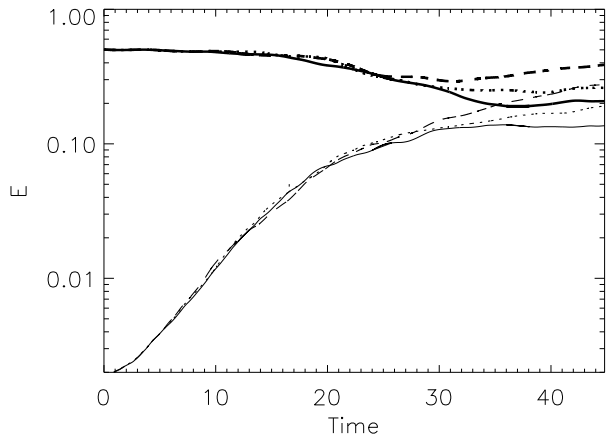} \end{center}
\caption{\label{fig:ener} Kinetic (thick curves) and magnetic 
    energy (thin curves) as a function of time, for runs 
    with $\epsilon = 0$ (solid lines), $\epsilon = 0.05$ (dotted lines), 
    and $\epsilon = 0.1$ (dashed lines).}
\end{figure}

In Hall-MHD, the Courant-Friedrichs-Levy (CFL) condition is more 
stringent than for MHD for which, with equipartition of kinetic and 
magnetic energy, the CFL condition for explicit time-stepping imposes 
an upper boundary on the time step $\Delta t \lesssim \Delta x/U_A$ 
where $\Delta x$ is the spatial step. In Hall-MHD, the dispersive 
nature of the whistlers impose 
$\Delta t \lesssim {\Delta x}^2/(\epsilon U_A)$. As a result, smaller 
time steps will be needed as $\epsilon$ is increased. Also, since the 
time step decreases quadratically as the spatial resolution is linearly 
increased, we cannot achieve now spatial resolutions higher than 
$256^3$ because of these constraints.

A helical forcing at $k_0=2$ given by an ABC flow 
{\setlength\arraycolsep{2pt}
\begin{eqnarray}
{\bf f} &=& \left[C\sin(k_0z)+B\cos(k_0y)\right] \hat{x} + 
    \left[A\sin(k_0x)+C\cos(k_0z)\right] \hat{y} + {} 
 \nonumber \\
&&{} + \left[B\sin(k_0y)+A\cos(k_0x)\right] \hat{z} ,
\end{eqnarray}}
with $A=0.9$, $B=1$, and $C=1.1$ was applied in the momentum equation. 
This election of the amplitude coefficients was done to ensure breaking 
the symmetries of the ABC flow and ensuring a faster development of 
turbulence \citep{Archontis03}. After a first hydrodynamic run made 
to reach a turbulent steady state, a random and small magnetic field was 
introduced at small scales. Initially the ratio of kinetic to magnetic 
energy was $E_u/E_b \sim 10^{-3}$. 

\begin{figure}
\begin{center} \includegraphics[width=9cm]{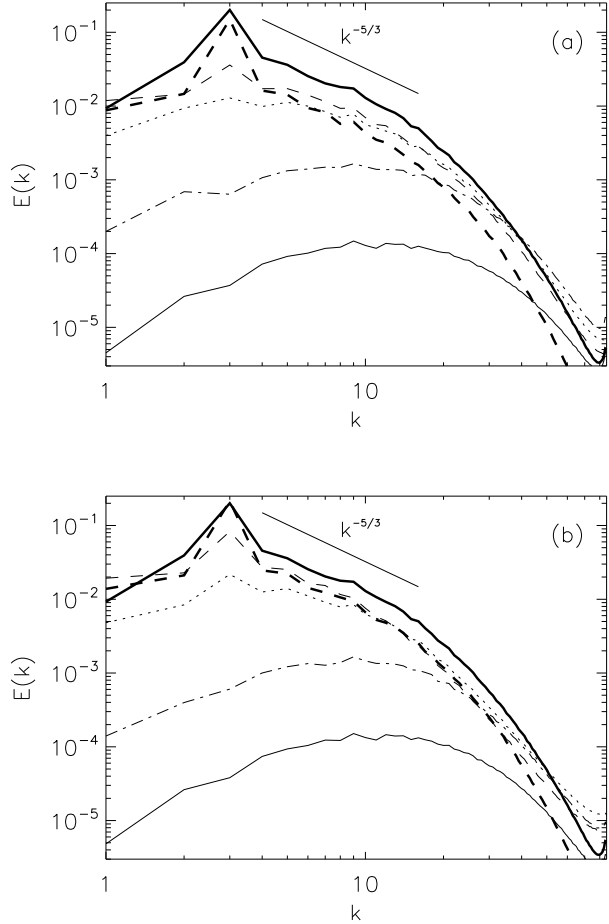} \end{center}
\caption{\label{fig:spec} Spectra of kinetic energy (thick curves) 
    and magnetic energy (thin curves) as a function of time, for 
    $t=5$ (solid), 15 (dash-dotted), 30 (dotted), and 45 (dashed). (a) 
    $\epsilon = 0.05$, and (b) $\epsilon = 0.1$.}
\end{figure}

The simulation was continued to see exponential growth of the magnetic 
energy (in the following, we will refer to this stage as the kinematic 
regime), and finally nonlinear saturation of the small scale magnetic 
field (in the following, Hall-MHD turbulence). Three simulations were 
done, with $\epsilon = 0$ (MHD), $\epsilon = 0.05$ (which corresponds to 
$k_{Hall} \approx 20$), and $\epsilon = 0.1$ ($k_{Hall} \approx 10$). 
Figure \ref{fig:ener} shows the time history of the kinetic and magnetic 
energies for these three runs. 

After $t \approx 20$, the small scale magnetic fields have reached 
saturation for all values of $\epsilon$, while the large scale magnetic 
field keeps growing slowly. As $\epsilon$ is increased, the magnetic 
energy reached by the system after the non-linear saturation of the 
small scales increases. However, this behavior is not monotonical in 
$\epsilon$ as shown by \citet{Mininni03b}.

\begin{figure}
\begin{center} \includegraphics[width=9cm]{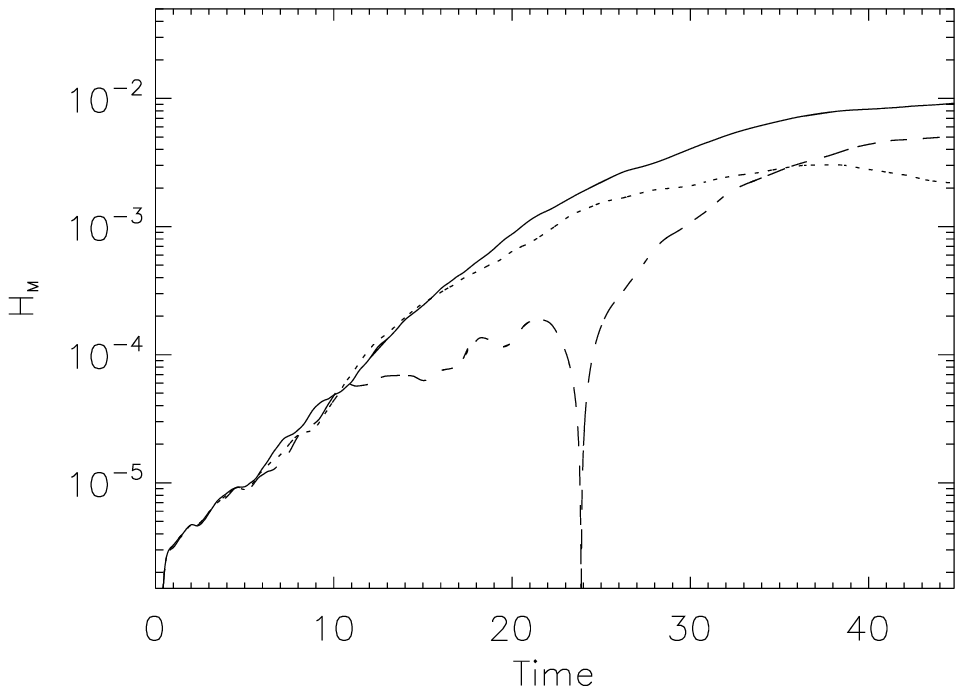} \end{center}
\caption{\label{fig:heli} Absolute value of the magnetic helicity as a 
    function of time. Labels are as in Fig. \ref{fig:ener}. For 
    $\epsilon=0.1$ the magnetic helicity changes sign from negative 
    to positive at $t\approx24$.}
\end{figure}

The saturation of the large scale magnetic field takes place in a longer 
time \citep{Brandenburg01,Mininni05a}. Note that one of the biggest 
challenges for DNS is to attain scale separation between the
different dynamical ranges that must be resolved. Reynolds numbers in 
simulations are much smaller than the values observed in astrophysics 
and geophysics. Moreover, compared with hydrodynamics and MHD, the extra 
characteristic length scale in Hall-MHD (the Hall scale) makes it even 
harder to achieve a proper separation between all these scales. As a 
result, we will focus in this work in the energy transfer at scales 
smaller than $k_0$ (the energy injection band), and the late time 
large-scale evolution of these runs will not be discussed here 
\citep[more details can be found e.g. in][]{Mininni05a}.

Figure \ref{fig:spec} shows the time evolution of the magnetic and 
kinetic energy spectra, for the runs with $\epsilon = 0.05$ and 
$\epsilon = 0.1$. As previously mentioned ($t \approx 40$) the 
spectrum of energy at scales smaller than $k_0$ has saturated and 
reached a steady state, while the magnetic energy at $k=1$ keeps 
growing slowly. Note that the ratio of kinetic to magnetic energy at 
small scales in the saturated state depends on the value of $\epsilon$.

Another quantity that will be of interest in the next section is the 
magnetic helicity. Figure \ref{fig:heli} shows the time history of 
the absolute value of magnetic helicity for the three runs. Note that 
while in the MHD run ($\epsilon = 0$) the magnetic helicity grows 
monotonically with time, in the Hall-MHD runs the time evolution is 
strongly modified. For $\epsilon = 0.05$ the magnetic helicity grows 
slower than in the MHD case, and for $\epsilon = 0.1$ it changes sign 
at $t \approx 24$. As was observed by \citet{Mininni03b}, the Hall 
effect inhibits the generation of net magnetic helicity at large scales 
by the helical dynamo process. This inhibition grows monotonically with 
the amplitude of Hall term, and for values of $\epsilon$ large enough 
the magnetic helicity fluctuates around zero. The reason for this 
behavior will be discussed in the next section.

\section{ \label{Transfers} Transfers }

In this section we discuss the energy transfer terms defined in 
Sec. \ref{Equations} as obtained from the three DNS discussed in 
the previous section.

\subsection{The run with $\epsilon = 0.1$}

\begin{figure}
\begin{center} \includegraphics[width=9cm]{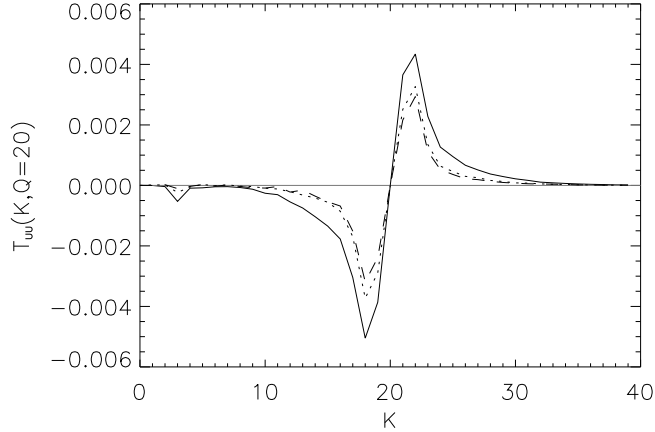} \end{center}
\caption{\label{fig:T1uu} Transfer of kinetic energy from $Q=20$ to $K$, 
    ${\mathcal T}_{UU}(K,Q=20)$, in the kinematic regime (solid line), 
    at $t=26$ (dotted line), and at $t=45$ (dashed line) in the run with 
    $\epsilon = 0.1$.}
\end{figure}

We start discussing in detail the transfer in the Hall-MHD run with 
$\epsilon =0.1$. At late times in this simulation, when the system 
is close to equipartition ($U_A \approx U$), the Hall wavenumber is 
$k_{Hall} \approx 10$. Since we consider transfer functions at different 
times, for the sake of comparison and unless explicitly said, all 
transfers in this subsection will be normalized using the {\it  r.m.s.} 
velocity and magnetic field according to their expressions 
[Eqs. (\ref{eq:TUU}-\ref{eq:TBBHall})]. Note that since $\epsilon {\bf J}$ 
has units of velocity (and ${\bf U}-\epsilon {\bf J}$ is the electron 
velocity), the transfer function $T_{BB}^\textrm{Hall}$ is normalized 
using $\left<B^2 |{\bf U}|\right>$. This election also allows for a 
direct comparison of this term against $T_{BB}^\textrm{MHD}$ [see Eq. 
(\ref{eq:TBBMHD})].

Figure \ref{fig:T1uu} shows the transfer of kinetic energy from 
the shell $Q=20$ to kinetic energy in shells $K$ for three different 
times. As previously mentioned, positive transfer denotes energy 
given by the shell $Q$, while negative transfer corresponds to 
energy received by this shell. In this case, kinetic energy 
in the shell $Q=20$ is mostly received from the shell $K=18$ 
(negative peak), and given to $K=22$ (positive peak). This 
function represents the local and direct transfer of kinetic energy 
to small scales. There are no noticeable differences in this transfer 
between the Hall-MHD runs ($\epsilon = 0.05$ and $0.1$), and the MHD 
run ($\epsilon = 0$).

The curve for early times (kinematic regime) corresponds to the 
initial exponential growth of magnetic energy, and is a time average 
properly normalized. As time evolves and the magnetic energy grows, 
the amount of kinetic energy transfered to small scales diminishes, 
since a larger amount of kinetic energy at large scales is turned 
into magnetic energy. This effect was previously observed in MHD runs 
\citep{Mininni05b}.

\begin{figure}
\begin{center} \includegraphics[width=9cm]{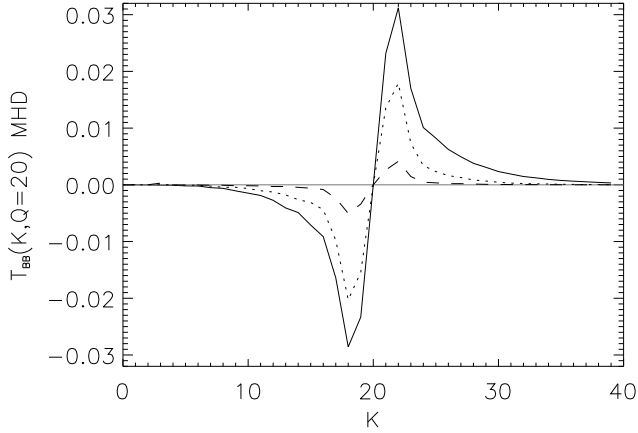} \end{center}
\caption{\label{fig:T1bbM} ${\mathcal T}_{BB}^\textrm{MHD}(K,Q=20)$ 
    in the kinematic regime (solid line), at $t=26$ (dotted line), and at 
    $t=45$ (dashed line) in the run with $\epsilon = 0.1$.}
\end{figure}

Figure \ref{fig:T1bbM} shows 
${\mathcal T}_{BB}^\textrm{MHD}(K,Q=20)$, the transfer of magnetic 
energy at the shell $Q=20$ to magnetic energy in shells $K$ due to the 
advection by the bulk velocity field. As in the case of 
${\mathcal T}_{UU}$, the transfer is local and the shell $Q=20$ 
receives most of the energy from $K=18$ (negative peak) and gives 
energy to the shell $K=22$ (positive peak). Again, no significant 
differences are observed between the three runs with different values 
of $\epsilon$, except that this transfer, in amplitude, gets 
substantially stronger as $\epsilon$ (and ${\bf B}$) increases.

\begin{figure}
\begin{center} \includegraphics[width=9cm]{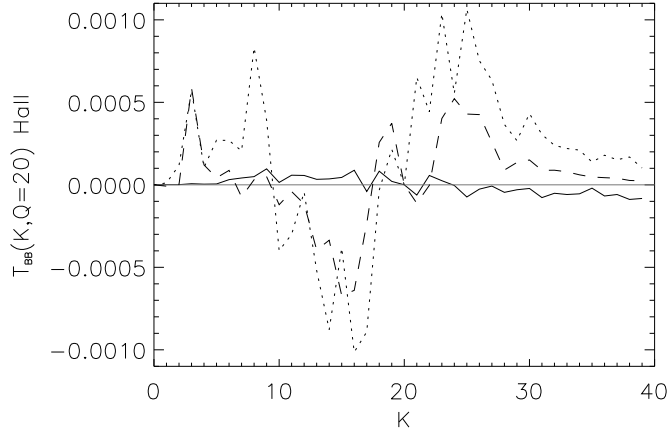} \end{center}
\caption{\label{fig:T1bbh} ${\mathcal T}_{BB}^\textrm{Hall}(K,Q=20)$ in 
    the kinematic regime (solid line), at $t=26$ (dotted line), and at 
    $t=45$ (dashed line) in the run with $\epsilon = 0.1$.}
\end{figure}

\begin{figure}
\begin{center} \includegraphics[width=9cm]{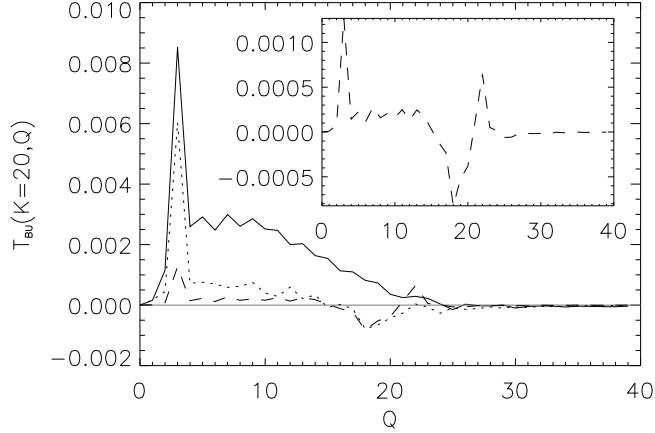} \end{center}
\caption{\label{fig:T1buk} ${\mathcal T}_{BU}(K=20,Q)$ in the kinematic 
    regime (solid line), at $t=26$ (dotted line), and at $t=45$ (dashed 
    line) in the run with $\epsilon = 0.1$.
    The inset shows a blow up of the last transfer.}
\end{figure}

The total shell-to-shell transfer of magnetic energy is given by 
${\mathcal T}_{BB}^\textrm{MHD}$ plus 
${\mathcal T}_{BB}^\textrm{Hall}$. Figure \ref{fig:T1bbh} shows the 
${\mathcal T}_{BB}^\textrm{Hall}(K,Q)$ transfer at $Q=20$. As in the 
previous cases, positive transfer denotes energy is given from the 
shell $Q=20$ to shells $K$, while negative transfer indicates the 
shell $Q$ receives energy from $K$. The 
${\mathcal T}_{BB}^\textrm{Hall}$ transfer is small during the kinematic 
regime, but grows as the small scales reach nonlinear saturation. 
Although this transfer is noisier than the previous terms studied, 
two regions can be identified at late times. Around $Q=K=20$, the 
transfer is local and direct: positive and negative peaks can be 
observed at $K \approx 25$ and $K \approx 15$, indicating 
energy is received and given respectively by the shell $Q$ from 
and to these wavenumbers. On the other hand, at large scales 
(up to $K \approx 10$) a region with positive transfer can also be 
identified. This region indicates a non-local and inverse transfer 
of energy: the shells with $K$ between 1 and 10 receive magnetic 
energy from the shell $Q=20$. This combination of a local direct 
transfer of energy and a non-local inverse transfer is characteristic 
of the Hall term, and is in qualitative agreement with the turbulent 
dissipation derived in Sec. \ref{Transport} where it was shown that it
can take negative values.

\begin{figure}
\begin{center} \includegraphics[width=9cm]{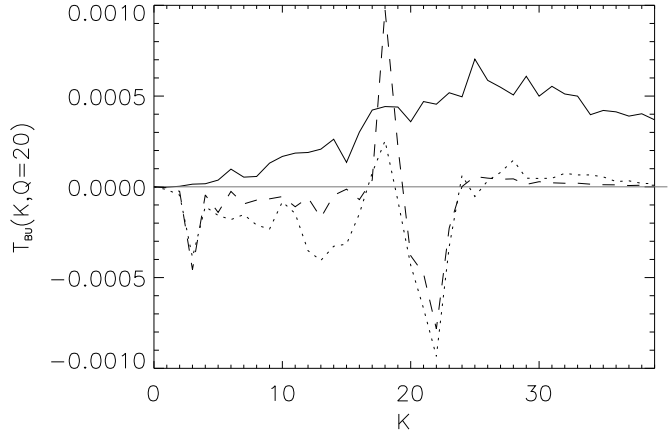} \end{center}
\caption{\label{fig:T1buq} ${\mathcal T}_{BU}(K,Q=20)$ at three 
    different times: the kinematic regime (solid line), $t=26$ (dotted 
    line), and  $t=45$ (dashed line) for the run with $\epsilon = 0.1$.}
\end{figure}

The remaining transfer term is ${\mathcal T}_{BU}(K,Q)$, which 
when positive represents transfer of kinetic energy from the shell 
$Q$ to magnetic energy in the shell $K$.  Although the expression 
of this transfer function is equal for MHD and Hall-MHD, the 
transfer is modified by the Hall currents. The reason for this can 
be explained in two ways. On the one hand, the expression of the 
$\alpha$-effect in Sec. \ref{Transport} is modified by the Hall term, 
and this term represents transfer of energy from the turbulent 
velocity field to the mean magnetic field. On the other hand, 
waves are expected to give non-local coupling between the velocity 
and magnetic fields \citep[see e.g.][in MHD]{Iroshnikov63,Kraichnan65}. 
In Hall-MHD, the non-dispersive Alfv\'en waves of MHD are replaced by 
dispersive circularly polarized waves and as a result, the coupling 
between the two fields should also be modified.

Figure \ref{fig:T1buk} shows ${\mathcal T}_{BU}(K=20,Q)$, the 
energy transfered to the magnetic field at $K=20$ from the velocity 
field at shells $Q$. In the kinematic regime this transfer is non-local 
and similar to the MHD transfer \citep{Alexakis05a,Mininni05b}: the 
magnetic field at $K=20$ receives energy from the large scale flow at 
$Q=3$ and from all turbulent scales up to $Q \approx 20$. However, at 
late times the transfer is strongly modified. The magnetic field at 
$K=20$ still receives energy from a broad range of wavenumbers $Q$ 
smaller than $K$ \citep[as was found in][]{Alexakis05a}, 
but it also receives energy from larger wavenumbers 
($Q \approx 22$), and gives energy to the velocity field at slightly 
smaller wavenumbers ($Q \approx 18$). Note that this indicates that in 
Hall-MHD a magnetic field at a given scale can give rise to velocity 
fluctuations at larger scales, a process studied by \citet{Mahajan05b} 
and referred there as the {\it reverse dynamo}.

\begin{figure*}
\begin{center} \includegraphics[width=12cm]{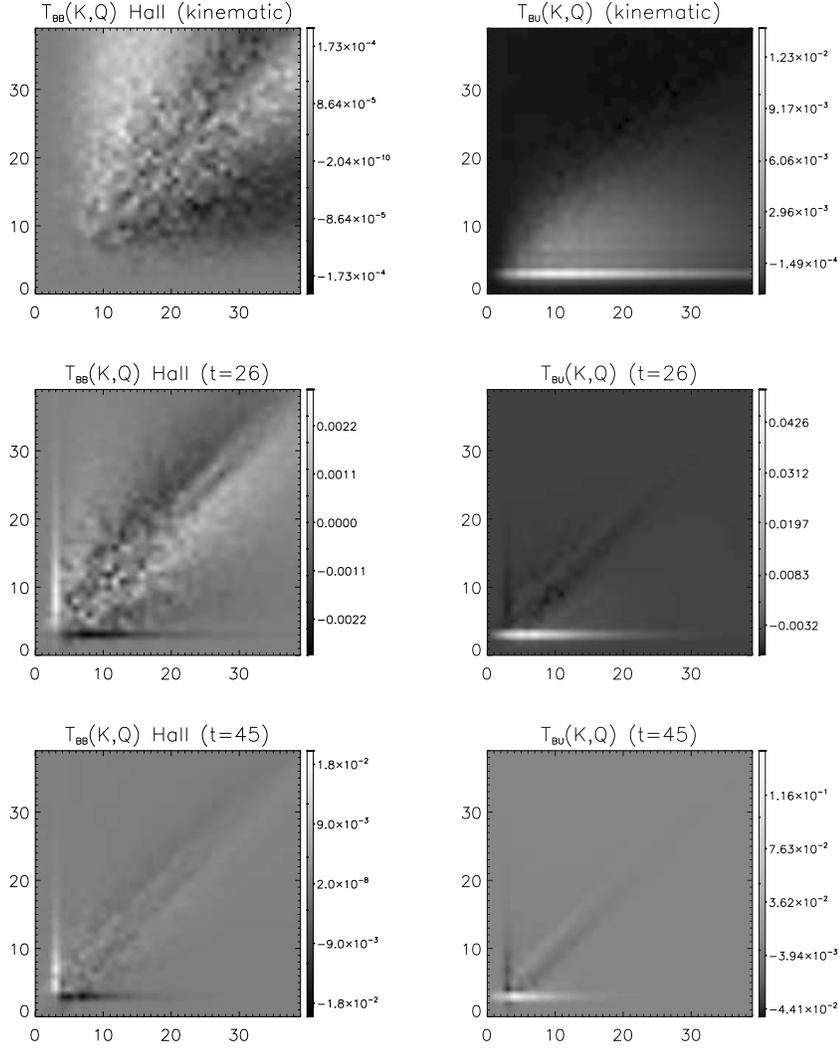} \end{center}
\caption{\label{fig:THall} ${\mathcal T}_{BB}^\textrm{Hall}(K,Q)$ 
    (left column), and ${\mathcal T}_{BU}(K,Q)$ (right column) at three 
    different times: the kinematic regime (top), $t=26$ (middle), and 
    $t=45$ (bottom) for the run with $\epsilon = 0.1$. In all figures, 
    $K$ is on the $x$ axis and $Q$ in the $y$ axis. Shading go from dark 
    (${\mathcal T}<0$) to light (${\mathcal T}>0$).}
\end{figure*}

Figure \ref{fig:T1buq} shows ${\mathcal T}_{BU}(K,Q=20)$, the 
energy received by the magnetic field at all wavenumbers $K$ 
from the velocity field in the shell $Q=20$.  During the kinematic regime, 
the velocity field in this shell gives energy to all magnetic shells, 
although the transfer peaks at wavenumbers larger than $Q$. But in 
the saturated regime, the transfer changes drastically again. The magnetic 
field at wavenumbers $K$ smaller than $Q \approx 16$, and in shells 
between 20 and 23 gives energy to the velocity field (negative transfer), 
while the magnetic field in shells between $K \approx 16 $ to 20 and for 
$K \gtrsim 23$ receives energy from the velocity field (positive transfer). 
This is just the counterpart of ${\mathcal T}_{BU}(K,Q)$ for constant $K$, 
and again shows that in Hall-MHD a small scale magnetic field can create 
large scale flows.

Figure \ref{fig:THall} shows shaded plots of 
${\mathcal T}_{BB}^\textrm{Hall}(K,Q)$ and ${\mathcal T}_{BU}(K,Q)$ 
at different times. These are the two transfers that are strongly modified 
by the Hall currents, and the figures allow for a study of the terms for all 
values of $K$ and $Q$. 
Although noisy, a characteristic pattern can be recognized in 
${\mathcal T}_{BB}^\textrm{Hall}$. As time evolves and the magnetic 
energy grows, the relative importance of this term grows. For 
wavenumbers $K,Q \gtrsim k_{Hall}\sim 10$, the function is positive 
(light) near and below the diagonal $K=Q$, and negative (dark) near and 
above this diagonal. This region close to the diagonal represents local 
and direct transfer of energy: a cut at constant $Q$ shows that close to 
the diagonal the shell $Q$ receives energy from neighboring shells with 
$K \lesssim Q$ (negative ${\mathcal T}_{BB}^\textrm{Hall}$) and gives 
energy to neighbor shells with $K \gtrsim Q$ (positive 
${\mathcal T}_{BB}^\textrm{Hall}$). As we move far from this diagonal, 
the sign of the regions above and below the diagonal changes. This 
indicates a non-local and inverse transfer of magnetic energy, from 
small to large scales, in agreement with the expression for the 
turbulent magnetic diffusivity obtained in Sec. \ref{Transport}.

The ${\mathcal T}_{BU}(K,Q)$ also shows an interesting behavior as a 
function of time. During the kinematic regime, ${\mathcal T}_{BU}$ is 
positive in a triangle defined by $K \gtrsim Q$. This indicates that the 
velocity field in a given shell amplifies the magnetic field in that shell 
and all the shells with larger wavenumber (smaller scales). Also a strong 
band around $Q = 3$ is observed, indicating that the velocity field in the 
energy injection band gives a lot of energy to the magnetic field. These 
results are similar to the kinematic MHD dynamo \citep[see][]{Mininni05b}. 
However, at late times an inverse process can be identified close to the 
diagonal $K=Q$. Above it, ${\mathcal T}_{BU}$ is positive, while below 
it, it is negative. This represents transfer of magnetic energy from a shell 
$K$ to kinetic energy in slightly smaller wavenumbers $Q$.

\begin{figure}
\begin{center} \includegraphics[width=9cm]{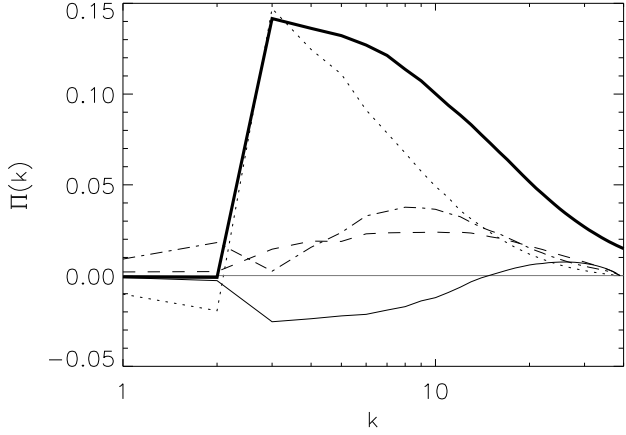} \end{center}
\caption{\label{fig:T1flu} Energy fluxes in the run with $\epsilon = 0.1$ at 
    $t=45$: $\Pi_{BB}^\textrm{Hall}(k)$ (solid line), 
    $\Pi_{BB}^\textrm{MHD}(k)$ (dash-dotted line), $\Pi_{UU}(k)$ (dashed), 
    and $\Pi_{BU}(k)$ (dotted). The thick line is the total flux 
    in the simulation.}
\end{figure}

Since ${\mathcal T}_{BB}^\textrm{Hall}$ and ${\mathcal T}_{BU}$ give both 
direct and inverse transfers of energy (locally or non-locally), it is of 
interest to quantify which direction wins when all the contributions to the 
transfer are added. To this end, we computed the contribution of each 
transfer term to the energy flux. The total energy flux at a wavenumber 
$k$ is given by
\begin{equation}
\Pi(k) = \sum_{K=0}^k \sum_Q {\mathcal T}(K,Q) ,
\end{equation}
where ${\mathcal T} = {\mathcal T}_{UU} + {\mathcal T}_{BB} + 
    {\mathcal T}_{UB} + {\mathcal T}_{BU}$ is the total energy transfer. We 
can split this flux into the energy flux due solely to the transfer of kinetic 
energy
\begin{equation}
\Pi_{UU}(k) = \sum_{K=0}^k \sum_Q {\mathcal T}_{UU}(K,Q) ,
\end{equation}
the flux due to the transfer of magnetic energy 
$\Pi_{BB} = \Pi_{BB}^\textrm{MHD} + \Pi_{BB}^\textrm{Hall}$, where
\begin{equation}
\Pi_{BB}^\textrm{MHD}(k) = \sum_{K=0}^k \sum_Q 
    {\mathcal T}_{BB}^\textrm{MHD}(K,Q) ,
\end{equation}
\begin{equation}
\Pi_{BB}^\textrm{Hall}(k) = \sum_{K=0}^k \sum_Q 
    {\mathcal T}_{BB}^\textrm{Hall}(K,Q) ,
\end{equation}
and the hybrid flux due to interactions between the velocity and magnetic 
fields
\begin{equation}
\Pi_{BU}(k) = \sum_{K=0}^k \sum_Q \left[ {\mathcal T}_{BU}(K,Q) + 
    {\mathcal T}_{UB}(K,Q) \right] .
\end{equation}
To compute the fluxes, the transfer functions are not normalized.

Figure \ref{fig:T1flu} shows the partial energy fluxes at $t=45$. Since 
all transfer functions were computed up to $K,Q=40$, the partial fluxes 
go to zero artificially at this wavenumber, although in the simulation 
the total energy flux goes to zero only at the maximum resolved wavenumber 
$k_{max}$. 

The total flux is positive at wavenumbers larger than $k_0$ (the energy 
injection band), indicating a direct cascade of the total energy. At 
wavenumbers smaller than $k_0$ the total flux is negative, an evidence 
of large scale dynamo action. A substantial portion of the total flux is due 
to the transfer of energy from the kinetic to the magnetic reservoirs 
($\Pi_{BU}$), and this contribution to the flux is positive at all 
wavenumbers (larger than the forced ones) indicating a net direct 
transfer of the energy. We note that this flux is due to the non-local 
$T_{UB}$ and $T_{BU}$ transfer terms. The flux due to the transfer of 
kinetic energy $\Pi_{UU}$ is also positive at all wavenumbers. But the 
flux due to the transfer of magnetic energy $\Pi_{BB}^{Hall}$ is only 
positive at wavenumbers larger than $k_{Hall}$. For wavenumbers smaller 
than $k \approx 10\sim k_{Hall}$, $\Pi_{BB}^\textrm{Hall}$ changes sign, 
giving as a result a net inverse transfer of magnetic energy, from small 
to large scales. This indicates that a magnetically dominated Hall-MHD 
system could display backscatter of the magnetic energy. Magnetic 
fluctuations at small scales could give rise to large scale magnetic 
fields, as is also implied by the expression of $\beta$ found in Sec. 
\ref{Transport}.

Note that although in MHD the inverse cascade of magnetic helicity 
can give a similar result, the backscatter predicted in 
Hall-MHD by the turbulent diffusivity is novel, since it can take place even 
in the absence of helicity in the fields. To illustrate this we show 
results of non-helical magnetically dominated simulations in Sec. 
\ref{Upscaling}. It is worth noting that at wavenumbers smaller than 
$k_{Hall}$, the Hall term increases the flux of magnetic energy to 
smaller scales, thus also in agreement with results showing the Hall 
currents increase the amount of small scale perturbations 
\citep{Birn01,Laveder02a,Laveder02b,Morales05}.

\subsection{Dependence with $\epsilon$}

Now we discuss in detail the dependence of the results as $\epsilon$ (or 
the Hall scale) is varied. To this end, we consider the runs with 
$\epsilon = 0.1$, $0.05$, and 0 (MHD). As previously mentioned, the 
transfer terms ${\mathcal T}_{UU}(K,Q)$ and ${\mathcal T}_{BB}(K,Q)$ do not 
show a dependence with the amplitude of the Hall effect. These terms give 
direct and local transfer of energy to small scales, as in MHD 
\citep{Alexakis05a}. As a result, we will discuss the change in the 
remaining transfer terms as $\epsilon$ is varied.

Since the Hall effect is more relevant when the magnetic field is stronger, 
we will consider the transfer terms at $t=45$, when a large scale magnetic 
field is present and the small scales have reached saturation. Since we 
will examine runs with different values of $\epsilon$ at the same time, 
in this subsection the transfers are not normalized using the energies. 

The transfer of magnetic energy due to the Hall term 
${\mathcal T}_{BB}^\textrm{Hall}(K,Q=20)$ is of course zero in the MHD 
case ($\epsilon=0$). As $\epsilon$ is increased, except for an increase 
in its amplitude (not shown), no significant differences are observed 
and its behavior is similar to the one examined in Sec. \ref{Transfers}.

\begin{figure}
\begin{center} \includegraphics[width=9cm]{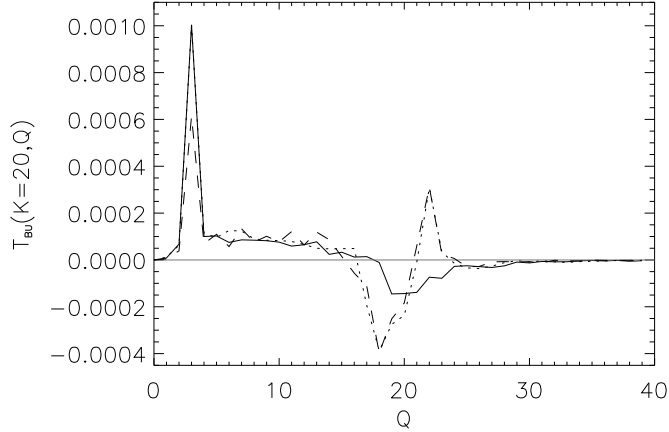} \end{center}
\caption{\label{fig:Tebuq} ${\mathcal T}_{BU}(K=20,Q)$ at $t=45$ for 
    $\epsilon = 0$ (solid line), $0.05$ (dotted line), and $0.1$ 
    (dashed line).} 
\end{figure}

Figure \ref{fig:Tebuq} shows the behavior of ${\mathcal T}_{BU}(K=20,Q)$ 
(the transfer of kinetic energy in the shells $Q$ to magnetic energy 
in the shell $K=20$) as $\epsilon$ is varied. The strong peak at $Q=3$ is 
associated with the injection band. This transfer is non-local in the three 
runs, as is evidenced by the positive plateau from $Q\approx3$ 
to $Q\approx 16$. As a result, the velocity field in all these shells 
gives energy to the magnetic field at $K=20$. As $\epsilon$ is increased, 
a local transfer grows in the neighborhood of $Q=20$. The velocity field 
at wavenumbers $K$ slightly larger give energy to the magnetic field at 
$K=20$, while the magnetic field gives energy to the velocity field at 
wavenumbers slightly smaller ($Q \approx 18$).

\begin{figure}
\begin{center} \includegraphics[width=9cm]{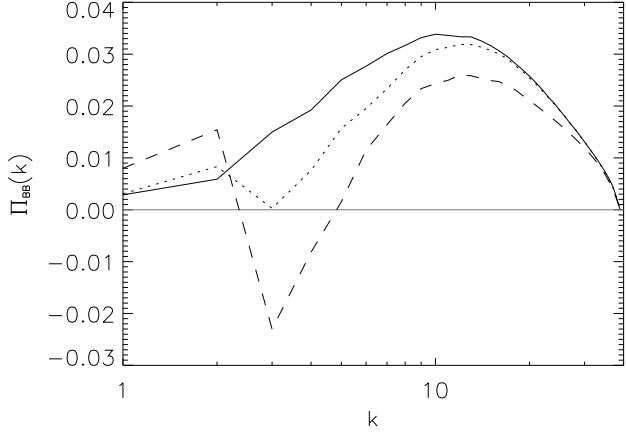} \end{center}
\caption{\label{fig:fluxbb} $\Pi_{BB}(k)$ at $t=45$ for $\epsilon = 0$ 
    (solid line), $0.05$ (dotted line), and $0.1$ (dashed line).}
\end{figure}

We can compute the energy flux as $\epsilon$ increases. Since the transfer 
of kinetic energy is not changed, we will focus on two contributions to 
the total flux: the flux of magnetic energy $\Pi_{BB}(k)$, and the 
hybrid flux $\Pi_{BU}(k)$ due to the terms turning kinetic into magnetic 
energy and vice-versa. The magnetic energy flux 
$\Pi_{BB} = \Pi_{BB}^\textrm{MHD} + \Pi_{BB}^\textrm{Hall}$ is shown in 
Fig. \ref{fig:fluxbb}. At scales larger than $k_{Hall}$, negative flux 
of magnetic energy is observed, giving backscatter of magnetic energy 
to large scales. As $\epsilon$ is increased, the amplitude of the 
backscatter grows, and the wavenumber where the flux changes sign moves 
to larger $k$.

\begin{figure}
\begin{center} \includegraphics[width=9cm]{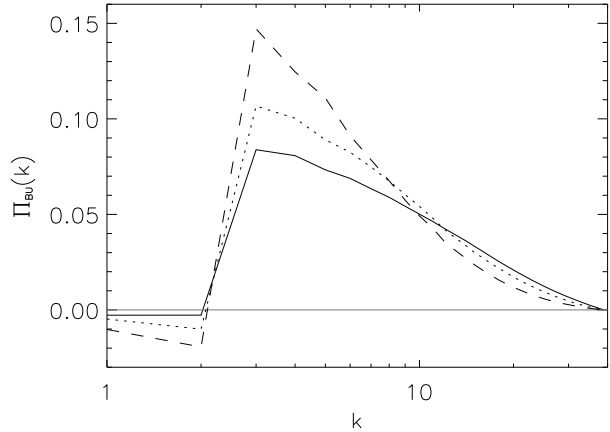} \end{center}
\caption{\label{fig:fluxbu} $\Pi_{BU}(k)$ at $t=45$ for $\epsilon = 0$ 
    (solid line), $0.05$ (dotted line), and $0.1$ (dashed line).}
\end{figure}

Figure \ref{fig:fluxbu} shows the flux $\Pi_{BU}(k)$. At wavenumbers 
smaller than the forcing wavenumber ($k=3$), the flux is negative. This 
is a signature of large scale dynamo action: the magnetic field at 
large scales is fed by the small scale velocity field. Remarkably, as 
$\epsilon$ is increased, the amplitude of the negative flux at large 
scales increases. This is in good agreement with dynamo simulations 
where the large scale magnetic field was observed to grow faster in the 
presence of Hall currents \citep{Mininni03b,Mininni05a}.

\subsection{Transfer of magnetic helicity}

We discuss briefly the transfer of magnetic helicity in the saturated 
case $t=45$. To study the transfer associated with the inverse cascade 
of magnetic helicity at scales larger than the forcing scale, a large 
separation between this scale and the largest scale in the box is needed. 
At a fixed spatial resolution, this reduces the Reynolds numbers, and 
as a result reduces also the separation between the Ohmic scale and the 
Hall scale. This study is beyond the aim of this work. But we want to 
point out a remarkable feature observed in the transfer of magnetic 
helicity at scales smaller than the forcing scale.

\begin{figure}
\begin{center} \includegraphics[width=9cm]{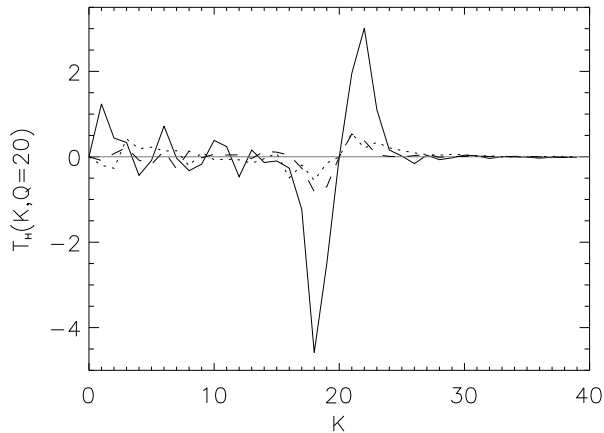} \end{center}
\caption{\label{fig:Thelicity} ${\mathcal T}_H(K,Q=20)$ normalized by the 
    magnetic helicity at the shell $Q=20$, at $t=45$ for $\epsilon = 0$ 
    (solid line), $0.05$ (dotted line), and $0.1$ (dashed line).}
\end{figure}

Figure \ref{fig:Thelicity} shows the transfer ${\mathcal T}_H(K,Q=20)$ 
normalized by the magnetic helicity in the shell $Q=20$ at $t=45$. 
The transfer is mostly local in the three simulations, peaking at 
wavenumbers $K$ slightly smaller and larger than $Q$. However, as 
$\epsilon$ is increased the transfer rate of magnetic helicity 
is strongly quenched. This slow down in the transfer in Hall-MHD 
explains the behavior observed in Fig. \ref{fig:heli}. In MHD and 
Hall-MHD dynamos, the external mechanical forcing generates equal 
amounts of magnetic helicity of opposite sign at scales smaller 
and larger than the forcing band \citep{Seehafer96,Brandenburg01,Mininni03b}. 
Since the transfer of magnetic helicity between different shells in 
the Hall-MHD runs is almost stopped, it takes more time for the 
magnetic helicity at scales smaller than the forcing scale to reach 
the dissipative scale where it can be destroyed. As a result, both 
signs of magnetic helicity piles up close to the forcing band, 
decreasing the growth rate of net magnetic helicity at scales larger 
than the forcing scale, and allowing also for 
the possibility for a sign change of the net magnetic helicity.

\section{ \label{Upscaling} Backscatter of magnetic energy in Hall-MHD}

The mechanically forced runs discussed in the previous section show 
negative flux of magnetic energy at large scales due to the Hall effect, 
in agreement with negative values of the turbulent diffusivity. This 
indicates that in a magnetically dominated simulation, backscatter of 
magnetic energy could be observed if the Hall term is strong enough. 
Note that here we are using the word {\it backscatter} to refer to this 
transfer of magnetic energy from the small to the large scales. This is 
done in opposition to the usual terminology of inverse cascades, since 
we have been unable to identify any ideal invariant of the Hall-MHD 
equations cascading inversely with constant flux to the large scales.

To study this scenario, we did three simulations with $\epsilon = 0$, 
$0.2$, and $0.5$. The kinematic viscosity and magnetic diffusivity 
were $\nu = \eta = 3.5 \times 10^{-2}$, and the spatial resolution was 
$N^3 = 128^3$. The initial condition was ${\bf U} = {\bf B} = 0$. The 
system was forced with a non-helical and random electromotive force given 
by a superposition of harmonic modes at wavenumbers $k=9$ and 10. The 
phases of the force were changed with a correlation time of 
$\tau = 1.25 \times 10^{-2}$, and the time step was set to 
$\Delta t = 1.5 \times 10^{-3}$. Note that in the absence of magnetic 
helicity, no inverse cascade is expected in the MHD case 
\citep[see however][for cases where a large scale shear is present]{Lanotte99}.

\begin{figure}
\begin{center} \includegraphics[width=9cm]{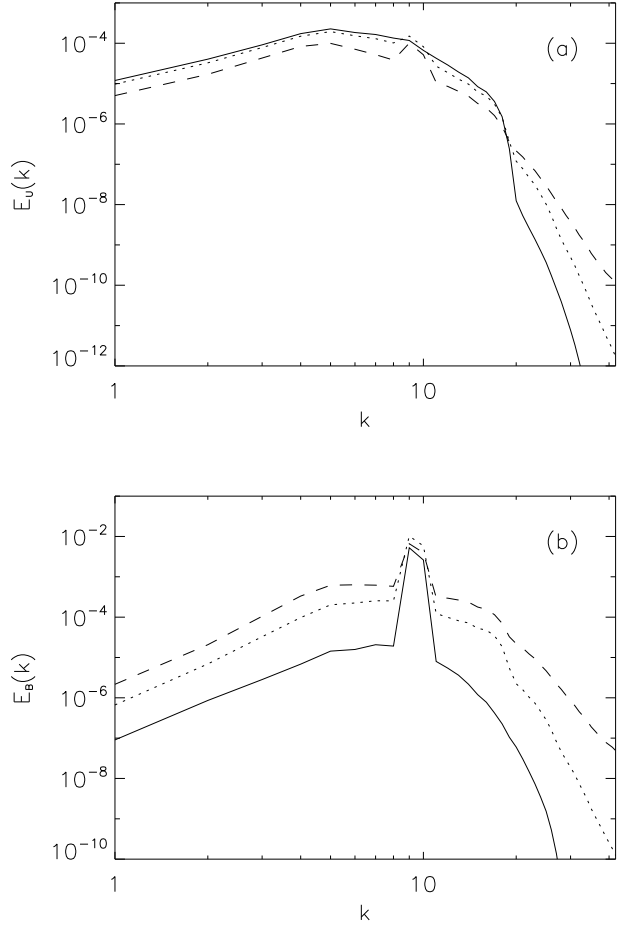} \end{center}
\caption{\label{fig:invsp1} (a) Kinetic energy spectrum $E_U(k)$ and 
    (b) magnetic energy spectrum $E_B(k)$ at $t=3$, for runs with 
    $\epsilon = 0$ (solid line), $0.05$ (dotted line), and $0.1$ 
    (dashed line).}
\end{figure}

\begin{figure}
\begin{center} \includegraphics[width=9cm]{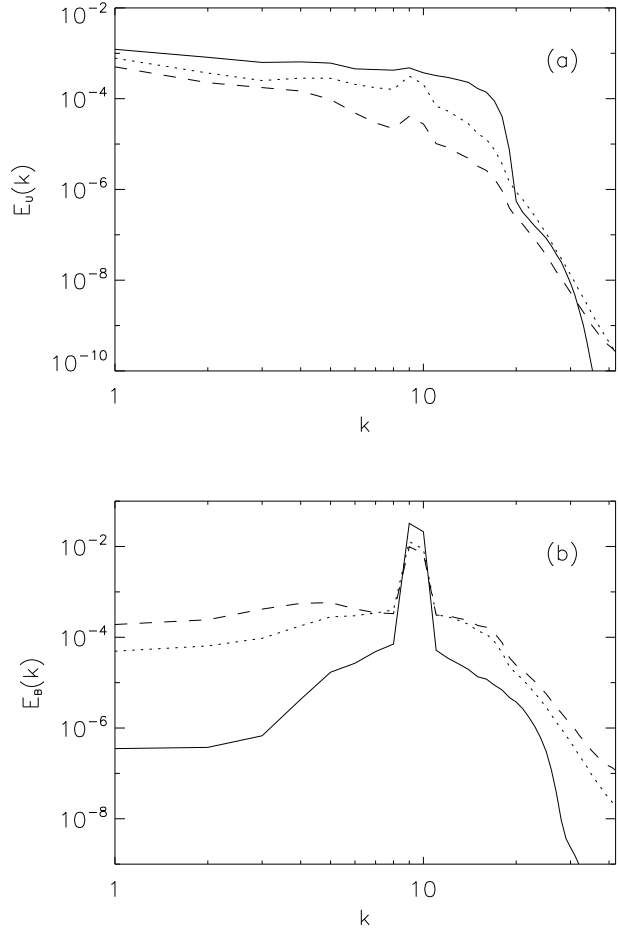} \end{center}
\caption{\label{fig:invsp2} (a) Kinetic energy spectrum and (b) 
    magnetic energy spectrum at $t=75$ for runs with $\epsilon = 0$ 
    (solid line), $0.05$ (dotted line), and $0.1$ (dashed line).}
\end{figure}

The system was run until reaching a turbulent steady state. All the 
quadratic invariants (with the exception of the total energy) were 
verified to be small: the magnetic helicity fluctuates in the three 
runs around zero, both the global quantity as well as its spectral 
density at each individual Fourier shell. Figure \ref{fig:invsp1} 
shows the kinetic and magnetic energy spectrum at early times ($t=3$). 
The shell of wavenumbers associated with the external magnetic force 
is easily recognized in the peak in Fig. \ref{fig:invsp1}(b).

As time evolves, an increase in the magnetic energy at wavenumbers 
smaller than the forcing wavenumber is observed. Figure \ref{fig:invsp2} 
shows the kinetic and magnetic energy spectrum at $t=75$, when the 
system has reached a steady state. The three runs are dominated by 
the magnetic energy (note that the peak in the magnetic energy spectrum 
around the forcing band gives the largest contribution to the energy). 
The spectrum of kinetic energy is similar for the three runs, and 
large scale perturbations are observed because of the injection of 
kinetic energy by the Lorentz force. However, the magnetic energy 
spectrum is strongly modified as $\epsilon$ is increased. While the 
spectra of the three simulations peak in the energy injection band, 
the magnetic energy at $K=1$ in the run with $\epsilon=0.5$ is 
three orders of magnitude larger than in the MHD run. The magnetic 
energy in all wavenumbers smaller than the forcing wavenumbers 
increases as $\epsilon$ is increased. 

The backscatter of magnetic energy is in good agreement with the 
negative flux $\Pi_{BB}$ observed in the previous section, and the 
negative turbulent transport coefficients derived for the magnetically 
dominated case. Note that an increase in the level of the small 
scale magnetic fluctuations (for wavenumbers smaller than the energy 
injection wavenumbers) is also observed in Fig. \ref{fig:invsp2}(b).

\section{ \label{Discussion} Discussion }

In this work we presented energy transfer in Hall-MHD turbulence as 
obtained from numerical simulations. The properties of the spectral 
transfer is one of the building blocks of turbulence theories, and to 
the best of our knowledge no attempt to study transfer and cascades of 
ideal invariants in this system of equations had been attempted before.

Before proceeding with the discussion of our results, we have to 
warn the reader about a clear limitation of the numerical results 
presented. As previously mentioned, an astrophysics-like scale 
separation between the box size, the energy injection scale, the Hall scale, 
and the Ohmic dissipation scale is well beyond today computing resources. 
We tested the dependence of our results as $\epsilon$ was varied, but 
no attempt was made to change the Reynolds numbers in our simulations. 
This being said, we believe that even under this limitation, an 
understanding of the transfer of energy between different scales is of 
uttermost importance for the development of a theory of turbulence for 
Hall-MHD or other extensions of magnetohydrodynamic to take into account 
kinetic plasma effects.

Direct evidence of nonlocality of the energy transfer was observed. While 
the total energy displays a direct cascade to small scales, in the 
individual transfer terms, both directions (toward small and large scales) 
were identified. Coupling between the magnetic and velocity fields is 
strongly modified by the Hall effect, and a local backscatter of energy 
from the magnetic field to the velocity field at slightly larger scales 
was observed. This behavior can be expected since the Hall term changes 
the nature of the nondispersive MHD Alfv\'en waves, into dispersive and 
circularly polarized waves. As a result, the nonlinear coupling between 
the two fields is also changed.

Also a nonlocal backscatter of magnetic energy was observed at scales 
larger than the Hall scale. This backscatter was verified in non-helical 
magnetically forced simulations, where the amplitude of the magnetic 
field at scales larger than the forcing scale was observed to grow in 
the Hall-MHD simulations, but not in the MHD run. In some sense, the 
magnetic field in Hall-MHD being frozen in the ideal case to the 
electron velocity field, couples non-locally both small scales 
(the current) and large scales (the bulk velocity field).

All these processes can be partially explained considering transport 
turbulent coefficients estimated from MFT. 
Unlike MHD, the turbulent diffusivity in Hall-MHD 
is not positive definite. In particular, its expression shows that 
ion-cyclotron waves are more likely to produce large values of negative 
(backscatter) or positive (reconnection) turbulent diffusivity than 
the whistler mode.

The transfer of magnetic helicity at small scales was also observed 
to be quenched by the Hall effect. While the mechanisms generating 
magnetic helicity in the Hall-MHD dynamo are the same as in MHD 
\citep{Mininni03b}, the transport of helicity is expected to be changed 
by the Hall currents \citep{Ji99}. As a result of the slow down in the 
transfer rate of magnetic helicity by the Hall effect, the late time 
evolution of the system is not characterized by a maximally helical 
large scale magnetic field as in the MHD case 
\citep{Pouquet76,Meneguzzi81,Brandenburg01}.

The Hall term gives a direct transfer of magnetic energy at scales 
smaller than the Hall scale, and an inverse transfer at scales larger 
than the Hall scale. This finding sheds light into the conflicting 
results reported in the literature, where the Hall effect was observed 
to increase the amount of small scales and magnetic dissipation in some 
cases, and to help large scale reorganization processes in other cases, 
as mentioned in the introduction.

As a result of this dual direction of the Hall transfer, a change in 
the power law followed by the total energy spectrum can be expected close 
to the Hall wavenumber. Steepening of the energy spectrum for wavenumbers 
smaller than $k_{Hall}$ was observed in 2.5D simulations with strong 
magnetic fields imposed, when the cross-correlation between the velocity 
and magnetic fields was significant \citep{Ghosh96}. In three dimensional 
dynamo simulations where the cross correlation is in general small, 
no change was observed \citep{Mininni05a}, although a faster 
growth of the large scale magnetic field was found. Given the nonlocal 
nature of the transfer in Hall-MHD, and the scale separation needed 
to observe a clear change in the energy spectrum, probably a huge 
increase in the spatial resolution is needed to confirm it.

\begin{acknowledgments}
Computer time was provided by NCAR. The NSF grant CMG-0327888
at NCAR supported this work in part and is gratefully acknowledged.
\end{acknowledgments}

\bibliography{ms}

\begin{thebibliography}{56}
\providecommand{\natexlab}[1]{#1}
\providecommand{\url}[1]{\texttt{#1}}
\providecommand{\urlprefix}{URL }

\bibitem[{Alexakis et~al.(2005{\natexlab{a}})Alexakis, Mininni and
  Pouquet}]{Alexakis05b}
Alexakis, A., Mininni, P.~D. and Pouquet, A. 2005{\natexlab{a}} \emph{Phys.\
  Rev.\ Lett.} Submitted.

\bibitem[{Alexakis et~al.(2005{\natexlab{b}})Alexakis, Mininni and
  Pouquet}]{Alexakis05a}
Alexakis, A., Mininni, P.~D. and Pouquet, A. 2005{\natexlab{b}} \emph{Phys.\
  Rev.\ E} In press.

\bibitem[{Archontis et~al.(2003)Archontis, Dorch and Nordlund}]{Archontis03}
Archontis, V., Dorch, S. B.~F. and Nordlund, A. 2003 \emph{Astron.\ Astrophys.}
  \textbf{410}, 759.

\bibitem[{Balbus and Terquem(2001)}]{Balbus01}
Balbus, S.~A. and Terquem, C. 2001 \emph{Astrophys.\ J.} \textbf{552}, 235.

\bibitem[{Bhattacharjee et~al.(1999)Bhattacharjee, Ma and
  Wang}]{Bhattacharjee99}
Bhattacharjee, A., Ma, Z.~W. and Wang, X. 1999 \emph{J.\ Geophys.\ Res.}
  \textbf{104}, 14543.

\bibitem[{Birn et~al.(2001)Birn, Drake, Shay, Rogers, Denton, Hesse,
  Kuznetsova, Ma, Bhattacharjee, Otto and Pritchett}]{Birn01}
Birn, J., Drake, J.~F., Shay, M.~A., Rogers, B.~N., Denton, R.~E., Hesse, M.,
  Kuznetsova, M., Ma, Z.~W., Bhattacharjee, A., Otto, A. and Pritchett, P.~L.
  2001 \emph{J.\ Geophys.\ Res.} \textbf{106}, 3715.

\bibitem[{Blackman and Field(1999)}]{Blackman99}
Blackman, E.~G. and Field, G.~B. 1999 \emph{Astrophys.\ J.} \textbf{521}, 597.

\bibitem[{Blackman and Field(2002)}]{Blackman02}
Blackman, E.~G. and Field, G.~B. 2002 \emph{Phys.\ Rev.\ Lett.} \textbf{89},
  265007.

\bibitem[{Brandenburg(2001)}]{Brandenburg01}
Brandenburg, A. 2001 \emph{Astrophys.\ J.} \textbf{550}, 824.

\bibitem[{Brandenburg and Subramanian(2005)}]{Brandenburg05}
Brandenburg, A. and Subramanian, K. 2005 \emph{Phys.\ Rep.} \textbf{417}, 1.

\bibitem[{Chen et~al.(2003{\natexlab{a}})Chen, Chen and Eyink}]{Chen03a}
Chen, Q., Chen, S. and Eyink, G.~L. 2003{\natexlab{a}} \emph{Phys.\ Fluids}
  \textbf{15}, 361.

\bibitem[{Chen et~al.(2003{\natexlab{b}})Chen, Chen, Eyink and Holm}]{Chen03b}
Chen, Q., Chen, S., Eyink, G.~L. and Holm, D.~D. 2003{\natexlab{b}}
  \emph{Phys.\ Rev.\ Lett.} \textbf{90}, 214503.

\bibitem[{Debliquy et~al.(2005)Debliquy, Verma and Carati}]{Debliquy05}
Debliquy, O., Verma, M.~K. and Carati, D. 2005 \emph{Phys.\ Plasmas}
  \textbf{12}, 042309.

\bibitem[{Ding et~al.(2004)Ding, Brower, Craig, Deng, Fiksel, Mirnov, Prager,
  Sarff and Svidzinski}]{Ding04}
Ding, W.~X., Brower, D.~L., Craig, D., Deng, B.~H., Fiksel, G., Mirnov, V.,
  Prager, S.~C., Sarff, J.~S. and Svidzinski, V. 2004 \emph{Phys.\ Rev.\ Lett.}
  \textbf{93}, 045002.

\bibitem[{Domaradzki and Rogallo(1990)}]{Domaradzki90}
Domaradzki, J.~A. and Rogallo, R.~S. 1990 \emph{Phys.\ Fluids} \textbf{2}, 413.

\bibitem[{Galanti et~al.(1995)Galanti, Kleeorin and Rogachevskii}]{Galanti95}
Galanti, B., Kleeorin, N. and Rogachevskii, I. 1995 \emph{Phys.\ Plasmas}
  \textbf{2}, 4161.

\bibitem[{Ghosh et~al.(1996)Ghosh, Siregar, Roberts and Goldstein}]{Ghosh96}
Ghosh, S., Siregar, E., Roberts, D.~A. and Goldstein, M.~L. 1996 \emph{J.\
  Geophys.\ Res.} \textbf{101}, 2493.

\bibitem[{Gruzinov and Diamond(1995)}]{Gruzinov95}
Gruzinov, A.~V. and Diamond, P.~H. 1995 \emph{Phys.\ Plasmas} \textbf{2}, 1941.

\bibitem[{Helmis(1968)}]{Helmis68}
Helmis, G. 1968 \emph{Mon.ber.\ dtsch.\ Akad.\ Wiss.\ Berlin} \textbf{10}, 280.

\bibitem[{Iroshnikov(1963)}]{Iroshnikov63}
Iroshnikov, P.~S. 1963 \emph{Sov.\ Astron.} \textbf{7}, 566.

\bibitem[{Ji(1999)}]{Ji99}
Ji, H. 1999 \emph{Phys.\ Rev.\ Lett.} \textbf{83}, 3198.

\bibitem[{Kraichnan(1965)}]{Kraichnan65}
Kraichnan, R.~H. 1965 \emph{Phys.\ Fluids} \textbf{8}, 1385.

\bibitem[{Krause and Raedler(1980)}]{Krause}
Krause, F. and Raedler, K.-H. 1980 \emph{Mean-field magnetohydrodynamics and
  dynamo theory}.
\newblock Pergamon Press, New York.

\bibitem[{Lanotte et~al.(1999)Lanotte, Noullez, Vergassola and
  Wirth}]{Lanotte99}
Lanotte, A., Noullez, A., Vergassola, M. and Wirth, A. 1999 \emph{Geophys.\
  Astrophys.\ Fluid Dyn.} \textbf{91}, 131.

\bibitem[{Laveder et~al.(2002{\natexlab{a}})Laveder, Passot and
  Sulem}]{Laveder02a}
Laveder, D., Passot, T. and Sulem, P.~L. 2002{\natexlab{a}} \emph{Phys.\
  Plasmas} \textbf{9}, 293.

\bibitem[{Laveder et~al.(2002{\natexlab{b}})Laveder, Passot and
  Sulem}]{Laveder02b}
Laveder, D., Passot, T. and Sulem, P.~L. 2002{\natexlab{b}} \emph{Phys.\
  Plasmas} \textbf{9}, 305.

\bibitem[{Lesieur(1997)}]{Lesieur}
Lesieur, M. 1997 \emph{Turbulence in fluids}.
\newblock Kluwer Academic Press.

\bibitem[{Mahajan et~al.(2005{\natexlab{a}})Mahajan, Mininni and
  G\'omez}]{Mininni05c}
Mahajan, S.~M., Mininni, P.~D. and G\'omez, D.~O. 2005{\natexlab{a}}
  \emph{Astrophys.\ J.} \textbf{619}, 1014.

\bibitem[{Mahajan et~al.(2005{\natexlab{b}})Mahajan, Shatasvili, Mikeladze and
  Sigua}]{Mahajan05b}
Mahajan, S.~M., Shatasvili, N.~L., Mikeladze, S.~V. and Sigua, K.~I.
  2005{\natexlab{b}} \emph{Astrophys.\ J.} Preprint doi:10.1086/'432867'.

\bibitem[{Mahajan and Yoshida(1998)}]{Mahajan98}
Mahajan, S.~M. and Yoshida, Z. 1998 \emph{Phys.\ Rev.\ Lett.} \textbf{81},
  4863.

\bibitem[{Meneguzzi et~al.(1981)Meneguzzi, Frisch and Pouquet}]{Meneguzzi81}
Meneguzzi, M., Frisch, U. and Pouquet, A. 1981 \emph{Phys.\ Rev.\ Lett.}
  \textbf{47}, 1060.

\bibitem[{Mininni et~al.(2005{\natexlab{a}})Mininni, Alexakis and
  Pouquet}]{Mininni05b}
Mininni, P.~D., Alexakis, A. and Pouquet, A. 2005{\natexlab{a}} \emph{Phys.\
  Rev.\ E} In press.

\bibitem[{Mininni et~al.(2002)Mininni, G\'omez and Mahajan}]{Mininni02}
Mininni, P.~D., G\'omez, D.~O. and Mahajan, S.~M. 2002 \emph{Astrophys.\ J.}
  \textbf{567}, L81.

\bibitem[{Mininni et~al.(2003{\natexlab{a}})Mininni, G\'omez and
  Mahajan}]{Mininni03b}
Mininni, P.~D., G\'omez, D.~O. and Mahajan, S.~M. 2003{\natexlab{a}}
  \emph{Astrophys.\ J.} \textbf{587}, 472.

\bibitem[{Mininni et~al.(2003{\natexlab{b}})Mininni, G\'omez and
  Mahajan}]{Mininni03a}
Mininni, P.~D., G\'omez, D.~O. and Mahajan, S.~M. 2003{\natexlab{b}}
  \emph{Astrophys.\ J.} \textbf{584}, 1120.

\bibitem[{Mininni et~al.(2005{\natexlab{b}})Mininni, G\'omez and
  Mahajan}]{Mininni05a}
Mininni, P.~D., G\'omez, D.~O. and Mahajan, S.~M. 2005{\natexlab{b}}
  \emph{Astrophys.\ J.} \textbf{619}, 1019.

\bibitem[{Mirnov et~al.(2003)Mirnov, Hegna and Prager}]{Mirnov03}
Mirnov, V.~V., Hegna, C.~C. and Prager, S.~C. 2003 \emph{Plasma Phys.\ Rep.}
  \textbf{29}, 566.

\bibitem[{Morales et~al.(2005)Morales, Dasso and G\'omez}]{Morales05}
Morales, L., Dasso, S. and G\'omez, D. 2005 \emph{J.\ Geophys.\ Res.}
  \textbf{110}, A04204.

\bibitem[{Numata et~al.(2004)Numata, Yoshida and Hayashi}]{Numata04}
Numata, R., Yoshida, Z. and Hayashi, T. 2004 \emph{Comp.\ Phys.\ Comm.}
  \textbf{164}, 291.

\bibitem[{Ohkitani and Kida(1992)}]{Ohkitani92}
Ohkitani, K. and Kida, S. 1992 \emph{Phys.\ Fluids A} \textbf{4}, 794.

\bibitem[{Ohsaki(2005)}]{Ohsaki05}
Ohsaki, S. 2005 \emph{Phys.\ Plasmas} \textbf{12}, 032306.

\bibitem[{Pouquet et~al.(1976)Pouquet, Frisch and L\'eorat}]{Pouquet76}
Pouquet, A., Frisch, U. and L\'eorat, J. 1976 \emph{J.\ Fluid Mech.}
  \textbf{77}, 321.

\bibitem[{Rezeau and Belmont(2001)}]{Rezeau01}
Rezeau, L. and Belmont, G. 2001 \emph{Space Sc.\ Rev.} \textbf{95}, 427.

\bibitem[{Rheinhardt and Geppert(2002)}]{Rheinhardt02}
Rheinhardt, M. and Geppert, U. 2002 \emph{Phys.\ Rev.\ Lett.} \textbf{88},
  101103.

\bibitem[{Sano and Stone(2002)}]{Sano02}
Sano, T. and Stone, J.~M. 2002 \emph{Astrophys.\ J.} \textbf{570}, 314.

\bibitem[{Seehafer(1996)}]{Seehafer96}
Seehafer, N. 1996 \emph{Phys.\ Rev.\ E} \textbf{53}, 1283.

\bibitem[{Shay et~al.(2001)Shay, Drake, Rogers and Denton}]{Shay01}
Shay, M.~A., Drake, J.~F., Rogers, B.~N. and Denton, R.~E. 2001 \emph{J.\
  Geophys.\ Res.} \textbf{106}, 3759.

\bibitem[{Smith et~al.(2004)Smith, Ghosh, Dmitruk and Matthaeus}]{Smith04}
Smith, D., Ghosh, S., Dmitruk, P. and Matthaeus, W.~H. 2004 \emph{J.\ Geophys.\
  Res.} \textbf{31}, L02805.

\bibitem[{Steenbeck et~al.(1966)Steenbeck, Krause and R{\"a}dler}]{Steenbeck66}
Steenbeck, M., Krause, F. and R{\"a}dler, K.-H. 1966 \emph{Z.\ Naturforsch.}
  \textbf{21a}, 369.

\bibitem[{Turner(1986)}]{Turner86}
Turner, L. 1986 \emph{IEEE Trans.\ Plasma Sci.} \textbf{PS-14}, 849.

\bibitem[{Verma(2004)}]{Verma04}
Verma, M. 2004 \emph{Phys.\ Rep.} \textbf{401}, 229.

\bibitem[{Waleffe(1991)}]{Waleffe91}
Waleffe, F. 1991 \emph{Phys.\ Fluids A} \textbf{4}, 350.

\bibitem[{Wang et~al.(2001)Wang, Bhattacharjee and Ma}]{Wang01}
Wang, X., Bhattacharjee, A. and Ma, Z.~W. 2001 \emph{Phys.\ Rev.\ Lett.}
  \textbf{87}, 265003.

\bibitem[{Yeung et~al.(1995)Yeung, Brasseur and Wang}]{Yeung95}
Yeung, P.~K., Brasseur, J. and Wang, Q. 1995 \emph{J. \ Fluid \ Mech.}
  \textbf{283}, 43.

\bibitem[{Zeldovich et~al.(1983)Zeldovich, Ruzmaikin and Sokoloff}]{Zeldovich}
Zeldovich, Y.~B., Ruzmaikin, A.~A. and Sokoloff, D.~D. 1983 \emph{Magnetic
  fields in astrophysics}.
\newblock Gordon and Breach Science Pub., New York.

\bibitem[{Zhou(1993)}]{Zhou93}
Zhou, Y. 1993 \emph{Phys.\ Fluids A} \textbf{5}, 2511.

\end{thebibliography}

\end{document}